\documentclass[prd,aps,floats,onecolumn,twoside,preprintnumbers,
superscriptaddress,floatfix,nofootinbib,showpacs]{revtex4}
\usepackage{graphicx,colordvi,pstricks,rotating,amsmath}
\usepackage{amssymb,epstopdf,bbm}
\usepackage{slashed}
\usepackage[T1]{fontenc}
\usepackage{epsfig}

\begin{document}
 
\title{Light-by-Light Scattering and Spacetime Noncommutativity}

\author{Raul Horvat}
\affiliation{Ru\dj er Bo\v{s}kovi\'{c} Institute, Bijeni\v{c}ka 54, 10000 Zagreb, Croatia}
\email{raul.horvat@irb.hr}
\author{Du\v sko Latas}
\affiliation{University of Belgrade, Faculty of Physics, P.O Box 44, Belgrade Serbia}
\email{latas@ipb.ac.rs}
\author{Josip Trampeti\'{c}}
\affiliation{Ru\dj er Bo\v{s}kovi\'{c} Institute, Bijeni\v{c}ka 54, 10000 Zagreb, Croatia}
\email{josip.trampetic@irb.hr}
\affiliation{Max-Planck-Institut f\"ur Physik, (Werner-Heisenberg-Institut), F\"ohringer Ring 6, D-80805 M\"unchen, Germany}
\email{trampeti@mppmu.mpg.de}
\author{Jiangyang You}
\affiliation{Ru\dj er Bo\v{s}kovi\'{c} Institute, Division of Physical Chemistry, Bijeni\v{c}ka 54, 10000 Zagreb, Croatia}
\email{jiangyang.you@irb.hr}
 
\newcommand{\tr}{\hbox{tr}}
\def\BOX{\mathord{\vbox{\hrule\hbox{\vrule\hskip 3pt\vbox{\vskip
3pt\vskip 3pt}\hskip 3pt\vrule}\hrule}\hskip 1pt}}
 
\date{\today}  

\begin{abstract}
The aim of this article is to explore a potential usability of a photon-photon self-interaction from the noncommutative quantum electrodynamics (NCQED) in the case of light-by-light scattering ($\gamma\gamma\to\gamma\gamma$) in ultraperipheral Pb+Pb collisions, a reaction  measured recently  by ATLAS and planned for future experiments in hadron-hadron colliders. We compute the total cross section from both, the full one-loop standard model (SM) and the tree-level NCQED amplitudes, in the equivalent photon approximation with impact parameters, for various noncommutative scales, $\rm\Lambda_{NC}$, and incoming nuclear spin-energy combinations. We find that NCQED contribution to the cross section has considerable increase at diphoton invariant mass range higher than  $\rm\Lambda_{NC}$, while the SM contribution is strongly suppressed in such region. Our results show that the current ATLAS $\rm\sqrt{s_{NN}} = 5.02$ TeV experiment can only probe $\rm\Lambda_{NC}<100$ GeV region. On the other hand, future hadron-hadron collider proposals could have the potential to extend to $\rm\Lambda_{NC}\lesssim 300$ GeV region, making the performance of $\rm Pb+Pb(\gamma\gamma)\to Pb+Pb\gamma\gamma$ scattering on testing space-time noncommutativity close to that of the previously proposed photon-photon mode of linear electron-positron collider.
\end{abstract}



\maketitle

Space-time noncommutativity and quantizations of the electromagnetic field in such noncommutative (NC) space, was an original idea to cure UV divergences plaguing quantum field theory, suggested by Heisenberg to Peierls, followed to Pauli, and finally executed a bit later by Oppenheimer's graduate student, Snyder \cite{Snyder:1946qz,Snyder:1947nq}.   Nowadays such models appear quite naturally in certain limits of String theory in the presence of a background $B^{\mu \nu}$ field. Specifically, a low-energy limit is identified where the entire boson-string dynamics in a Neveu-Schwartz condensate is described by a minimally coupled supersymmetric gauge theory on noncommutative space \cite{Seiberg:1999vs} such that the mathematical framework of NC geometry/field theory \cite{Madore:2000en,Jackiw:2001jb} does apply.

Of various types of noncommutative space-time structures the one generated by a constant real antisymmetric matrix $\theta^{\mu\nu}$ of dimension $length^2$, i.e. the Moyal space, has attracted a large amount of research interests for several crucial reasons: It emerges from string theory in a constant $B^{\mu \nu}$ field background~\cite{Seiberg:1999vs}. Pertubatively quantized gauge theories can be constructed through the nonlocal  Moyal-Weyl star($\star$)-product realization~\cite{Martin:1999aq}, and the existence of intriguing mathematical structure called Seiberg-Witten (SW) map~\cite{Seiberg:1999vs}. Thanking to these properties, NCQFT and related topics have been constantly investigated throughout time for over two decades now~\cite{Szabo:2009tn,Hinchliffe:2002km,Hewett:2000zp,Calmet:2001na,Behr:2002wx,Aschieri:2002mc,Schupp:2008fs,Horvat:2010sr,Trampetic:2015zma,Gomis:2000zz,Aharony:2000gz,Carroll:2001ws,Chaichian:2000si,Abel:2006wj,Horvat:2010km,Horvat:2011qn,Horvat:2017gfm,Horvat:2017aqf,Martin:1999aq,Hayakawa:1999yt,Hayakawa:1999zf,Martin:2016zon,Martin:2016hji,Martin:2016saw,Martin:2017nhg,Lust:2017wrl}. Particle physics models on NC space-time close to the existing standard model have been successfully in via NC gauge theories in Moyal space \cite{Hewett:2000zp,Calmet:2001na,Behr:2002wx,Aschieri:2002mc,Horvat:2011qn}. Various bounds on NC scales have been established over time through these models built \cite{Chaichian:2000si,Hewett:2000zp,Carroll:2001ws, Hinchliffe:2002km, Szabo:2009tn,Horvat:2010sr, Horvat:2010km, Horvat:2017gfm, Horvat:2017aqf}.

One direct consequence of noncommutativity is that gauge bosons (i.e. photons) in a NC U(1) gauge theory become self coupled. This property has very influential consequences in the $\theta$-exact NCQFT which are still under investigation now~\cite{Martin:1999aq,Gomis:2000zz, Aharony:2000gz, Schupp:2008fs, Trampetic:2015zma, Martin:2016zon, Martin:2016hji, Martin:2016saw, Martin:2017nhg}. NC photon self-coupling also induces nontrivial tree-level corrections to photon related processes like pair production, pair annihilation, Compton scattering, and most importantly, the tree-level photon-photon light-by-light (L-by-L) scattering process. The latter is considered a good candidate process in search for possible signal of space-time noncommutativity because the same process is suppressed at tree-level in standard model due to the  Landau-Young theorem.

Studies on light-by-light scattering used to focus on the photon-photon mode of linear electron-positron collider~\cite{Badelek:2001xb,Hewett:2000zp,Jikia:1993tc,Gounaris:1999gh,Bern:2001dg}. Recently a new method for observing light-by-light  scattering from the ultraperipheral heavy ion scattering has attracted much attention from society~\cite{Baur:2001jj,dEnterria:2013zqi,Klusek-Gawenda:2016euz,Harland-Lang:2018iur,Aaboud:2017bwk,Aad:2019ock,Ellis:2017edi,Akmansoy:2018xvd,Kostelecky:2018yfa}. In such scenario the electromagnetic field of high relativistic incoming heavy ions X (X=$^{208}$Pb, $^{197}$Au,....) is approximated as distribution of almost on-shell photons (equivalent photon approximation). In short, process  $\rm XX\to XX\gamma\gamma$ can then be treated as convolution of incoming photon distributions and $\gamma\gamma\to\gamma\gamma$ scattering~\cite{Baur:2001jj,dEnterria:2013zqi,Klusek-Gawenda:2016euz,Harland-Lang:2018iur,Aaboud:2017bwk,Aad:2019ock,Ellis:2017edi,Kostelecky:2018yfa}. Initial observation of such processes on the ATLAS experiment of LHC strongly boosts the interest on light-by-light scattering~\cite{Aaboud:2017bwk,Aad:2019ock}. Heavy ion collisions are also considered in the proposals for next generation circular hadron collider(s)~\cite{SppCPbPb,Armesto:2014iaa,Dainese:2016gch}.

Using peripheral scattering scenario as a test-bed for various beyond standard model theories have also been frequently considered. The theories engaged in testing the photon self-coupling in ultraperipheral $\rm PbPb\to PbPb\gamma\gamma$  collisions include perturbatively expanded Born-Infeld theory, some nonlinear corrections to Maxwell electrodynamics, and local Lorentz violating operators \cite{Ellis:2017edi,Akmansoy:2018xvd,Kostelecky:2018yfa}, but not the nonlocal Lorentz violating  operators in NCQED. In this work, we present a new computation for the tree level photon-photon scattering in LHC Pb-Pb collisions in the minimal $\theta$-exact SW mapped NCQED model \cite{Schupp:2008fs,Trampetic:2015zma}, in scenarios associated to this and next generation hadron collider(s) and estimate the capability of these scenarios in probing the NC scales.  

Under the assumption that main deviations from the SM in the exclusive  $\gamma\gamma\to\gamma\gamma$ scattering measurements are originating from the NCQED, in this letter we continue to search for the scale $\rm\Lambda_{NC}$ of the NC spacetime by using our $\theta$-exact minimal gauge sector action
\begin{equation}
S_{\rm gauge}^{\rm min}=\int-\frac{1}{2}F^{\mu\nu}\star F_{\mu\nu},
\;\;
F_{\mu\nu}=\partial_\mu A_\nu - \partial_\nu A_\mu -i(A_\mu\star A_\nu-A_\nu\star A_\mu), 
\label{NCminAction}
\end{equation}
given in terms of the NC gauge field $A_\mu$. The connection between the noncommutative and commutative fields is given by the $\theta$-exact Seiberg-Witten maps. Perturbative quantization is done by the standard BRST procedure~\cite{Martin:2016zon}. Note that our approach here does not involve an expansion over the deformation parameter $\theta$  itself. Therefore our results below do not have an upper energy limit by the noncommutative scale.

Through lengthy and straightforward diagramatic computation one can prove that the NC scattering amplitudes are invariant under the $\theta$-exact SW maps, i.e. all additional contributions induced in the action (\ref{NCminAction}) by the SW map cancel out \cite{LTY}.

The $\gamma\gamma\to\gamma\gamma$ scattering amplitudes in NCQED w.o. SW map can be solved from a  color -- $\star$-product equivalence between the ordinary U(N) and the Moyal NC U(1) gauge theories~\cite{Raju:2009yx, Huang:2010fc,LTY},
\begin{equation}
\tr \;\prod_{i=1}^n T^{\alpha_i}\; \Longleftrightarrow \;
\exp\Big({-\frac{i}{2} \sum_{i=1}^n k_{2i-1}\theta k_{2i}}\Big).
\label{Helicity1}
\end{equation}
By this means the noncommutative structure decouples with the $\star$-product ordered amplitudes, making the latter identical to the corresponding QCD color-ordered amplitudes.
After a summation over contributing color orders, we reach the L-by-L  helicity  scattering amplitudes of NCQED~\cite{Hewett:2000zp}. Of all possible helicity combinations only three are non-zero and independent~\cite{Hewett:2000zp,LTY}:
\begin{gather}
\begin{split}
M_{\rm NC}^{++++}&
=32\pi\alpha\Big(\frac{s}{u}\sin\frac{k_1\theta k_2}{2}\sin\frac{k_3\theta k_4}{2}-\frac{s^2}{tu}\sin\frac{k_1\theta k_4}{2}\sin\frac{k_2\theta k_3}{2}\Big),
\\
M_{\rm NC}^{+-+-}&
=32\pi\alpha\Big(\frac{t^2}{su}\sin\frac{k_1\theta k_2}{2}\sin\frac{k_3\theta k_4}{2}-\frac{t}{u}\sin\frac{k_1\theta k_4}{2}\sin\frac{k_2\theta k_3}{2}\Big),
\\
M_{\rm NC}^{++--}&
=32\pi\alpha\Big(\frac{u}{s}\sin\frac{k_1\theta k_2}{2}\sin\frac{k_3\theta k_4}{2}-\frac{u}{t}\sin\frac{k_1\theta k_4}{2}\sin\frac{k_2\theta k_3}{2}\Big).
\label{HelicityAmpl}
\end{split}
\end{gather}


We then evaluate the NCQED contribution to the ultraperipheral $\rm PbPb\to PbPb\gamma\gamma$ process in the equivalent photon approximation~\cite{Baur:2001jj,Klusek-Gawenda:2016euz}. To do this we sum over the tree-level NCQED and one-loop standard model amplitudes \cite{Jikia:1993tc,Gounaris:1999gh,Bern:2001dg} to evaluate the exclusive $\gamma\gamma\to \gamma\gamma$ differential cross in polar coordinates:
\begin{equation}
\begin{split}
&\frac{d\sigma(\omega_1,\omega_2,\vartheta,\varphi)}{d\Omega}=\frac{1}{(16\pi)^2}
\frac{1}{(\omega_1(1-\cos\vartheta)+\omega_2(1+\cos\vartheta))^2}
{\sum\limits_h} \big|M^h_{\rm SM} + M^h_{\rm NC}\big|^2,\;d\Omega=\sin\vartheta d\vartheta\, d\varphi.
\end{split}
\label{SMNCAmplitudeSquareHelicities}
\end{equation} 

Next we compute the total $\rm PbPb\to PbPb\gamma\gamma$ cross section by convoluting the exclusive differential cross section with the photon distribution of the equivalent photon beams, i.e. the photon number function $N(\omega_i, |\vec b_i|)$ ($i=1,2$). Here 2D vectors $\vec b_i$ are the impact parameters~\cite{Baur:2001jj,Klusek-Gawenda:2016euz} that marks the position of the ion from the position of impact in the plane perpendicular to the beam direction.  Peripheral nature of the collision requires $|\vec b_1-\vec b_2|>R_{min}=2R_{\rm N}$ to exclude direct collision between nuclei when integrating over beams. The incoming photon spectrum space  ($\omega_1,\omega_2$) is usually re-parametrized by the diphoton invariant mass $m_{\gamma\gamma}=\sqrt{4\omega_1\omega_2}$ and rapidity $Y=\frac{1}{2}\ln\frac{\omega_1}{\omega_2}$. Taking all these into account the cross section of $\rm PbPb\to PbPb\gamma\gamma$ is expressed in the fiducial phase space as an eightfold integral:
\begin{equation}
\sigma=\frac{1}{2}\int dm_{\gamma\gamma} \,dY  \frac{d^2 L_{\gamma\gamma}}
{d m_{\gamma\gamma}\,dY}d\Omega\frac{d\sigma(\omega_1,\omega_2,\vartheta,\varphi)}{d\Omega},  
\label{TotFidu}
\end{equation}
by convoluting (\ref{SMNCAmplitudeSquareHelicities}) with luminosity function $\frac{d^2L_{\gamma\gamma}}{dm_{\gamma\gamma}\,dY}$: 
\begin{equation}
\begin{split}
&\frac{d^2 L_{\gamma\gamma}}{dm_{\gamma\gamma}\, dY} =\frac{4\pi}{m_{\gamma\gamma}}\int\limits_0^\infty d |\vec b_1|\int\limits_0^\infty d |\vec b_2| \int\limits_0^{2\pi}d\phi N(\omega_1, |\vec b_1|)N(\omega_2, |\vec b_2|)\,
\Theta\left({\sqrt{|\vec b_1|^2+|\vec b_2|^2-2|\vec b_1||\vec b_2|\cos\phi}-R_{min}}\right),
\end{split}
\label{intb1b2}
\end{equation}
where we take $R_{min}=2R_{\rm Pb}=14$ $ \rm fm\simeq 71 \; GeV^{-1}$. In this work the  fourfold integration over impact parameters $\vec b_1$ and $\vec b_2$ is performed with the photon number function from monopole form factor~\cite{Baur:2001jj,Klusek-Gawenda:2016euz}
 \begin{equation}
N(\omega,b)=\frac{Z^2\alpha}{\pi^2}\left(\frac{\omega}{\gamma}K_1\left(\frac{\omega b}{\gamma}\right)-\sqrt{\frac{\omega^2}{\gamma^2}+\xi^2}\cdot K_1\left(b\sqrt{\frac{\omega^2}{\gamma^2}+\xi^2}\right)\right)^2, \;
\gamma=\frac{\sqrt{s_{\rm NN}}}{2 m_u}.
\end{equation}
In the above $K_1$ is the modified Bessel function of 2nd kind, $\xi=0.088$ GeV for $^{208}$Pb ion, $\gamma$ is the ion Lorenz factor and $ m_u=0.931\,{\rm GeV}$ is the atomic mass unit. 

Remaining integration over $dm_{\gamma\gamma}\,dYd\Omega$ has to be performed over an domain defined according to the experimental selection rules.  For a typical cut which requires transverse energy $E_T\ge E_0$ and both outgoing particles to bear absolute pseudorapidities $\left|y_{3,4}\right|\le y_0$, the full integral domain is defined as follows:
\begin{equation}
\begin{split}
&\varphi\in [0, 2\pi],\; |\cos\vartheta|\le\frac{e^{2y_0}-1}{e^{2y_0}+1},\; m_{\gamma\gamma}\ge 2E_0,
\\
Y\in & 
\left[\frac{1}{2}\left(\frac{1}{2}\ln\frac{1+\cos\vartheta}{1-\cos\vartheta}-y_0\right), \frac{1}{2}\left(y_0-\frac{1}{2}
\ln\frac{1+\cos\vartheta}{1-\cos\vartheta}\right)\right]
\\&
\cap\left[\frac{1}{2}\ln\frac{1+\cos\vartheta}{1-\cos\vartheta}+\ln\left(\frac{m_{\gamma\gamma}}{2E_0}-\sqrt{\frac{m_{\gamma\gamma}^2}{4E_0^2}-1}\right), 
\frac{1}{2}\ln\frac{1+\cos\vartheta}{1-\cos\vartheta}+\ln\left(\frac{m_{\gamma\gamma}}{2E_0}+\sqrt{\frac{m_{\gamma\gamma}^2}{4E_0^2}-1}\right)\right].
\end{split}
\label{Constraints}
\end{equation} 
From~\cite{Aaboud:2017bwk, Aad:2019ock} we have $E_0=3$ GeV and $y_0=2.37$ to define the ATLAS cuts. These cuts and  $\gamma=2693$ for ATLAS experiment with lead ion center of mass energy $\sqrt{s_{\rm NN}}=5.02$ TeV \cite{Aad:2019ock} allows us to estimate full SM one-loop contributions to the fiducial cross section: $\rm \sigma_{SM}(PbPb \to PbPb\gamma\gamma)=57\,nb$, comparable to the previously reported values of $\simeq$ 50 nb \cite{Aad:2019ock}.

\begin{figure}[t]
\begin{center}
\includegraphics[width=17cm,angle=0]{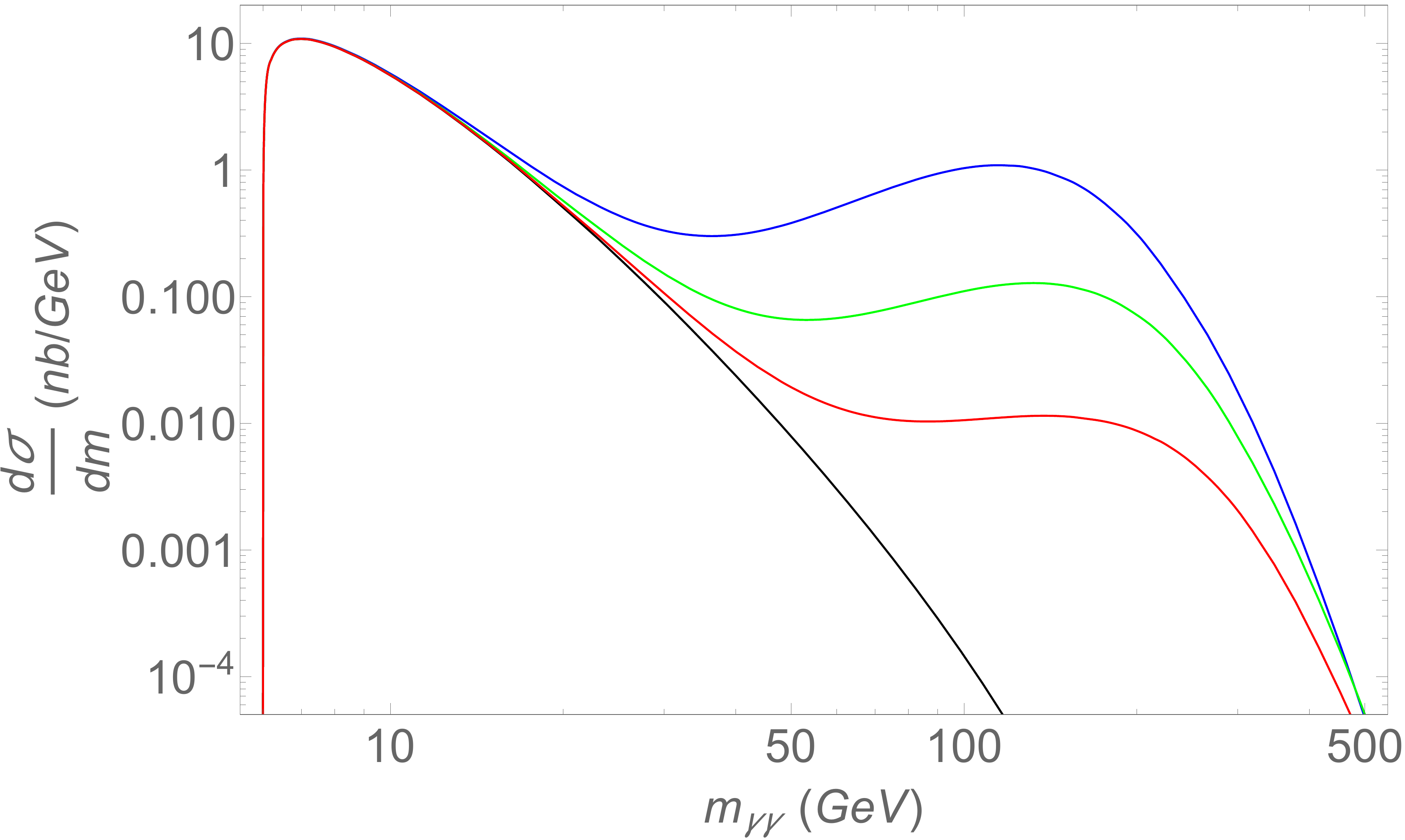}
\end{center}
\caption{Cross section versus diphoton invariant mass distribution in the $\rm PbPb \to PbPb\gamma\gamma$ experiment with the ATLAS cuts at $\sqrt{s_{\rm NN}}=5.02$ TeV, for the pure SM (black) as well as the SM+NCQED with $\rm\Lambda_{NC}$ values 53 (blue), 72 (green) and 100 (red) GeV.}
\label{figure1}
\end{figure}

We then numerically compute a combined contribution from the SM one-loop and the NCQED tree-level $\gamma\gamma\to\gamma\gamma$ amplitudes to the $\rm PbPb \to PbPb\gamma\gamma$ collision at several $\rm\Lambda_{NC}$ values. 
The NC tensor $\theta^{\mu\nu}$ is set to have only space-space components $\theta^{i3}\neq 0$ for simplicity. The third axis is set to be parallel to the beam, as usual. The result indicates that NC contribution takes place mainly at diphoton mass range considerably higher than the dominating part of the SM contribution (FIG.~\ref{figure1}) when $\rm\Lambda_{NC}$ is so valued that NC contribution to the cross section is comparable with that of SM (TABLE I).
We therefore consider that the high diphoton mass peak from the NCQED contribution determines the probability for the experiment to detect it. Our results presented in TABLE I and FIG.\ref{figure1} show that diphoton invariant mass distribution peak values of
  $\rm\sim(1,0.1,0.01)\,\frac{nb}{GeV}$ exist for $\rm\Lambda_{NC}\gtrsim (53,72,100)$ GeV, respectively.
Given the current integrated luminosity accumulation speed ($\rm\sim 1$ $\rm nb^{-1}$/yr) of current LHC and assuming that high luminosity (HL) LHC will reach about ten times of this value, we estimate that the current ATLAS experiment could only probe $\rm\Lambda_{NC}< 100$ GeV. 
\begin{table}
\begin{center}
\begin{tabular}{|c|c|c|c|c|c|c|}
\hline
$\rm\Lambda_{NC}$ (GeV) &  $\sigma_{\rm SM}$ (nb) & $\sigma_{\rm NC}$ (nb) & $\sigma_{\rm NC^2}$ (nb)  & $\sigma_{\rm NCQED}$ (nb) & $\rm\frac{d\sigma_{\rm NCQED}}{dm_{\gamma\gamma}}\big|_{max} \big( \frac{nb}{GeV}\big)$  & $\rm m_{\gamma\gamma}|_{max}$ (GeV) \\
\hline
53 & 57&12.1 & 125.5 & 137.2 & 1.09 & 115 \\
72 & 57&3.6 & 17.6 & 21.2 & 0.13 & 132 \\
100 & 57&1.0 & 1.8 & 2.8 & 0.011 & 138 \\
\hline
\end{tabular}
\caption{Summary of the predicted $\sigma_{\rm SM}$ and $\sigma_{\rm NCQED}=\sigma_{\rm NC}+\sigma_{\rm NC^2}$ contributions to $\rm PbPb \to PbPb\gamma\gamma$ fiducial cross sections for the ATLAS cuts at $\sqrt{s_{\rm NN}}=5.02$ TeV and various $\Lambda_{\rm NC}$ values. $\rm m_{\gamma\gamma}|_{max}$ denotes the diphoton mass position of $\rm\frac{d\sigma_{\rm NCQED}}{dm_{\gamma\gamma}}\big|_{max}$.}
\end{center}
\label{table2}
\end{table}
\begin{figure}[t]
\begin{center}
\includegraphics[width=17cm,angle=0]{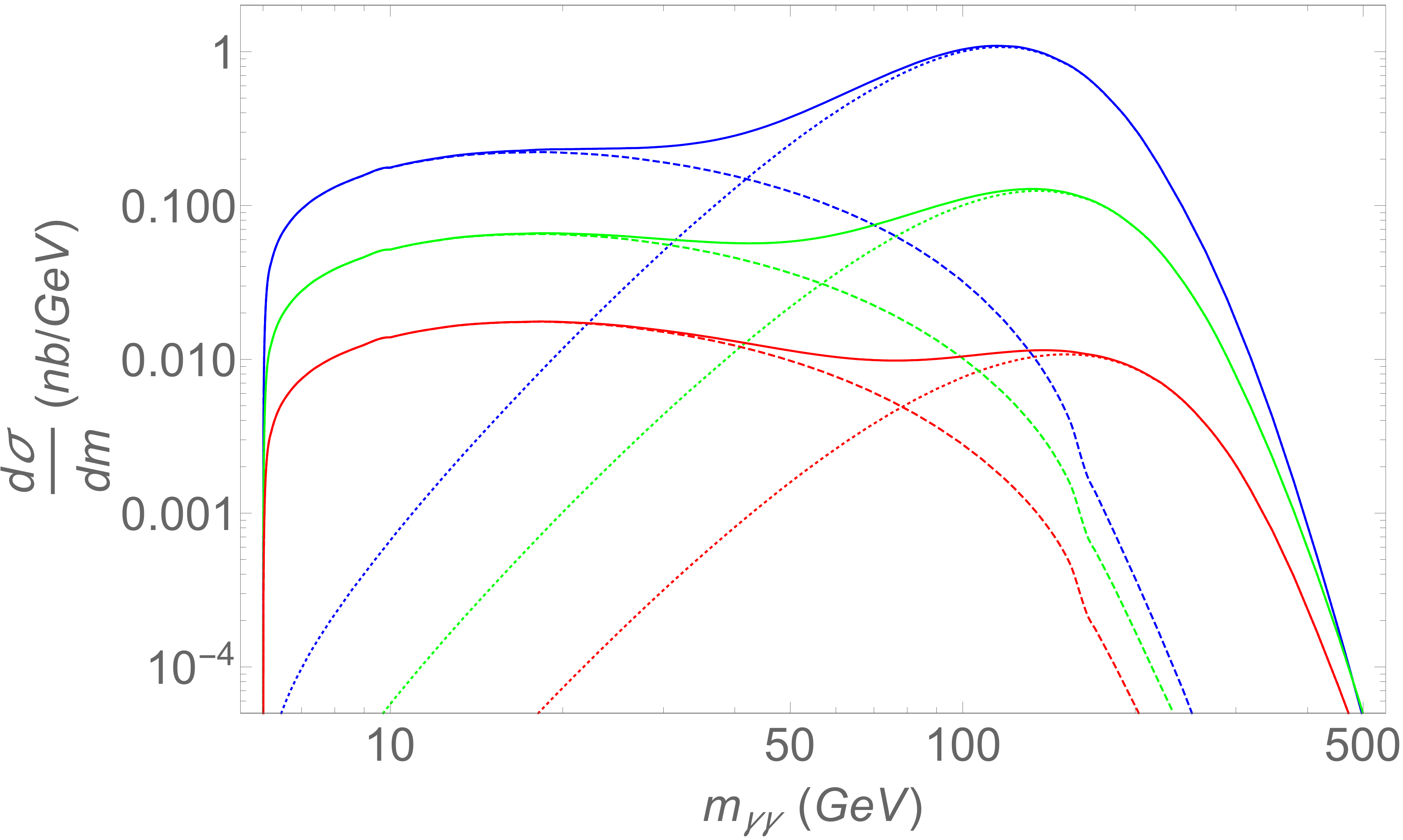}
\end{center}
\caption{NCQED related contributions to the cross section at $\sqrt{s_{\rm NN}}=5.02$ TeV versus diphoton invariant mass distributions under the ATLAS cuts conditions for $\Lambda_{\rm NC}$ values 53 (blue), 72 (green) and 100 (red) GeV. Dotted lines are pure noncommutative contribution ($\rm NC^2$), dashed lines are SM$\times$NCQED interference (NC), and solid lines are sum: $\rm NC+ NC^2=NCQED$.}
\label{figure2}
\end{figure}

The existence of a second peaks, with maxims at $\rm m_{\gamma\gamma}|_{max}$ reflect the evolution of noncommutative factors with respect to energy scales:  When energy scales are much smaller than $\rm\Lambda_{\rm NC}$, the NC factors are increasing as monomials with high power. Once the energy scales become larger than $\rm\Lambda_{\rm NC}$ the NC factors become oscillatory and bounded. Consequently the NC amplitudes are very small at very low energies where SM contribution dominates, and then increase fast and compete with the exponentially decreasing luminosity factor~\cite{Baur:2001jj} to give the rising side of the NC second peak. Once the energy scale goes beyond  $\rm\Lambda_{\rm NC}$ the NC amplitudes start to deviate from monomial increase, so that $\rm\frac{d\sigma}{dm_{\gamma\gamma}}$ falls down quickly. The subsequent oscillatory behaviors of the NC factors are fully suppressed by the luminosity function and can not be seen in this process. Consequently, the NC$^2$ contribution tends to reach a maximum at diphoton mass scales higher than corresponding NC contribution, as shown in FIG. 2. The latter also shows a kind of plateau in the plot $\rm\frac{d\sigma}{dm_{\gamma\gamma}}$ because of the interference pattern.  Matching the reported experimental total cross section of 78 nb~\cite{Aad:2019ock} by such combination yields a very small NC scale $\rm\Lambda_{NC}$ of about 72 GeV. 
This does not explain the experimental excess since the cross section induced by the NCQED amplitude mainly comes from the region $\rm m_{\gamma\gamma}\gtrsim 100$ GeV, which turns out to be  much above the experimentally measured excess of $\sim$ 30 GeV, as nicely presented by ATLAS Collaboration in FIG.2b of \cite{Aad:2019ock}.

Because of the limited probing capability of current ATLAS experiment, we move on to estimate potential improvements/enhancements one would expect from future generation circular collider(s). We consider proposals for next generation hadron collider Super Proton-Proton Collider (SppC) with energy $\sqrt{s_{\rm pp}}=$70 TeV \cite{Canbay:2017rbg}, and from the Future Circular Collider (FCC) proposal up to $\sqrt{s_{\rm pp}}=$100 TeV \cite{Acar:2016rde,Abada:2019zxq,Abada:2019lih,Benedikt:2018csr,Abada:2019ono}, as well as a far future scenario with $\sqrt{s_{\rm NN}}\simeq$100 TeV. We assume that all the ATLAS cuts kinematics remains the same, except the energy scale/Lorentz factor which scales up to 5, 7 or 20 times with respect to the current ATLAS value $\sqrt{s_{\rm NN}}=$5.02 TeV, and estimate the NC scale $\rm\Lambda_{NC}$ corresponding to a high energy $\rm\frac{d\sigma}{dm_{\gamma\gamma}}\big|_{max}$ with $\rm\sim 0.01$ nb/GeV magnitude. TABLE II shows that $\rm\Lambda_{NC}$ are about 2.5, 3.1 and 5.2 times of the $\rm\sim 100$ GeV limit for the ATLAS energy scale 5.02 TeV, respectively. Thus we conclude that future ultraperipheral heavy ion collision experiment(s) could probe $\rm\Lambda_{NC}$ scales close to L-by-L scattering in the photon-photon mode of a linear $\rm e^{-}e^{+}$ collider  ($\rm\sim 500$ GeV)~\cite{Hewett:2000zp}.
\begin{table}[t]
\begin{center}
\begin{tabular}{|c|c|c|c|c|c|c|c|}
\hline
$\rm\sqrt{s_{NN}}$ (TeV) & $\gamma$ & $\Lambda_{\rm NC}$ (GeV) & $\sigma_{\rm SM}$ (nb) & $\sigma_{\rm NCQED}$ (nb) & $\rm\frac{d\sigma_{\rm NCQED}}{dm_{\gamma\gamma}}\big|_{max} \big( \frac{nb}{GeV}\big)$&
$\rm m_{\gamma\gamma}|_{max}$ (GeV)
\\
\hline
5.02 & 2693 & 100 & 57 & 2.8 & 0.011 &  138
\\
25.10 & 13465 & 257 & 178 & 6.6 &  0.011 &  567
\\
35.14 & 18851 & 311 & 211 & 7.9 &  0.010 &  737
\\
100.40 & 53860 & 523 & 336 & 16.9 &  0.011 &  1480
\\
\hline
\end{tabular}
\caption{Estimations for ATLAS $\rm PbPb \to PbPb\gamma\gamma$ like experiments at higher energies. Here we made adjustments of the NC scale in a way to obtain $\rm\frac{d\sigma_{NCQED}}{dm_{\gamma\gamma}}\big|_{max}$ $\rm\simeq 0.01 \big( \frac{nb}{GeV}\big)$, at relevant $\rm m_{\gamma\gamma}|_{max}$ diphoton mass points given in this Table.}
\end{center}
\label{table3}
\end{table}

\begin{figure}[t]
\begin{center}
\includegraphics[width=17cm,angle=0]{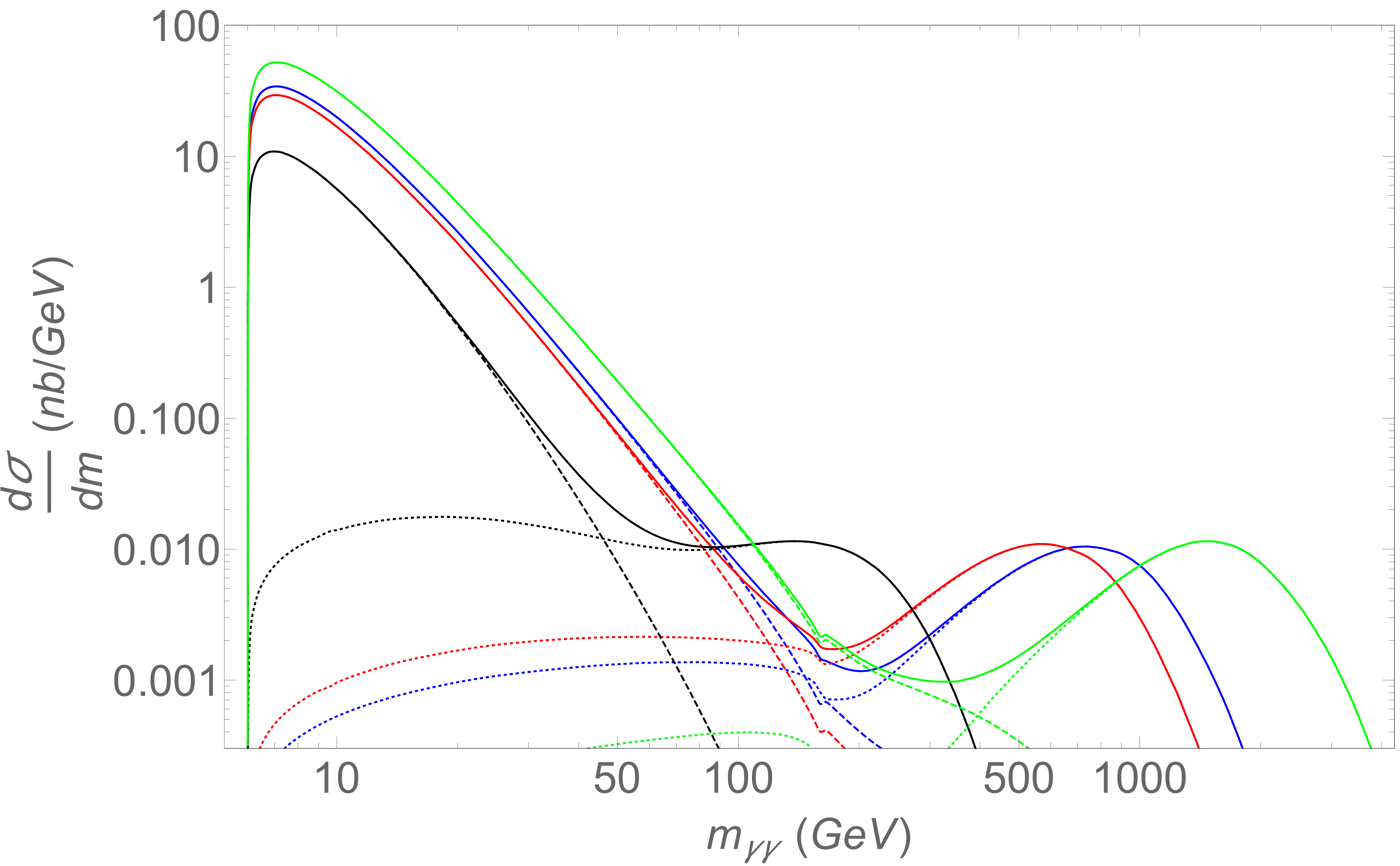} 
\end{center}
\caption{Cross section versus diphoton invariant mass distribution in the future higher energy ATLAS $\rm PbPb \to PbPb\gamma\gamma$ like experiments. Dashed curves are for SM contributions, dotted curves for NCQED, and solid correspond to the SM+NCQED, respectively. Black curves are for $\rm\sqrt{s_{NN}}=5.02$ TeV, $\rm\Lambda_{NC}=100$ GeV; red curves are for $\rm\sqrt{s_{NN}}=25.10$ TeV, $\rm\Lambda_{NC}=257$ GeV; blue curves are for $\rm\sqrt{s_{NN}}=35.14$ TeV, $\rm\Lambda_{NC}=311$ GeV, and green curves are for $\rm\sqrt{s_{NN}}=100.40$ TeV, $\rm\Lambda_{NC}=523$ GeV. }
\label{figure3}
\end{figure}

Our estimate show that like in the original ATLAS case (FIGs.~\ref{figure1} and~\ref{figure2}), NCQED contribution in the next generation hadron collider ultraperipheral heavy ion scattering scenarios also manifests as a second peak of $\rm\frac{d\sigma}{dm_{\gamma\gamma}}$ at diphoton mass range moderately higher than $\rm\Lambda_{NC}$ (FIG.~\ref{figure3}).  In addition FIG.~\ref{figure3} transparently shows that the point at which NCQED starts decoupling from the SM, and the NCQED contribution second peaks, as functions of ($\rm\sqrt{s_{NN}},\Lambda_{NC}$) follows more or less the similar  sort of pattern. One can also notice in FIG.~\ref{figure3} with increased collision energies, that the $\rm NC+NC^2$ plateau and $\rm NC^2$ peak becomes more separated than in FIG.~\ref{figure2}. Thus an small dent at $\rm m_{\gamma\gamma}\in [150, 200]\, GeV$ due to SM W-loops become visible in the total NCQED plots for next generation collider scenarios in  FIG.~\ref{figure3}, while only visible in the interference plots from FIG.~\ref{figure2}. Such effect is unfortunately experimentally invisible. 
Similar high diphoton mass peak appeared in perturbatively expanded Born-Infeld theory \cite{Ellis:2017edi}, too. However, it is to be mentioned that in the latter case the peak actually locates in diphoton mass range close to the deformation parameter. Therefore, one has to extrapolate out of the perturbative regime (as the authors of \cite{Ellis:2017edi} were aware themselves), to obtain credible results there. On the other hand our results, presented in this paper, stay free from such a restriction as a consequence of the $\theta$-exact approach.

The integrated luminosity ($\rm\gg 100$~$\rm nb^{-1}$) required for detecting the NCQED second peak at the scale $\rm\sim 0.01$ nb/GeV is difficult to achieve for both the present day LHC ($\rm> 1 nb^{-1}/yr$) and the future high luminosity (HL) LHC ($\rm>3 nb^{-1}/yr$)~\cite{Jowett:2015dmf,Jowett:2019jni}, respectively. However it would be attainable if the current performance projection ($\rm\sim 30 nb^{-1}/yr$) of next generation hadron collider could be reached/exceeded~\cite{Dainese:2016gch}. It is also worthy to note that at each $\rm m_{\gamma\gamma}|_{max}$'s from TABLEs I and II, in FIGs \ref{figure1} and  \ref{figure3}, the corresponding $\rm \frac{d\sigma_{SM}}{dm_{\gamma\gamma}}$ values are  several  orders of magnitude below the second peaks.

In summary, our calculations show that ultraperipheral $\rm PbPb \to PbPb\gamma\gamma$ scattering experiment on next generation hadron collider(s) could probe $\rm\Lambda_{NC}$ up to $\rm\sim 300$ GeV, if a high integrated luminosity $\rm\gg 100$~$\rm nb^{-1}$ and proper diphoton invariant mass range $\rm m_{\gamma\gamma}|_{max}\gtrsim 3\Lambda_{NC}$ can be achieved by the technical design. Such performance would be close to the light-by-light scattering in the photon-photon mode of a linear $\rm e^{-}e^{+}$ collider. We therefore consider ultraperipheral heavy ion scattering a potential future alternative to photon-photon (mode of  linear $\rm e^{-}e^{+}$) collider in probing space-time noncommutativity on accelerators and wish this perspective can be included in the technical design considerations of the next generation hadron collider(s). 

\noindent

\acknowledgments
J.T. would like to thank ATLAS  collaboration colleagues, M. Dyndal and M. Schott, for discussions regarding PbPb collision experiments at the LHC.  J.T. also acknowledge Wolfgang Hollik, Dieter Lust and Peter Minkowski for many discussions and to thank Max-Planck-Institute for Physics for hospitality. 
The work of R.H. and J.Y. has been supported by Croatian Science Foundation. The work of D.L. is supported by Serbian Ministry of Education, Science and Technological Development under the project No. ON171031. 
Computation was done by the package \textsc{LoopTool}~\cite{Hahn:1998yk} and \textsc{Mathematica} \cite{mathematica}.

\end{document}